\documentclass[twocolumn,showpacs,prl,10pt]{revtex4-1}
\usepackage[english]{babel}
\usepackage{amsmath,amssymb} 
\usepackage{times}
\usepackage{epsfig}
\begin{document} 

\title{Finite-temperature Drude weight within the anisotropic Heisenberg chain}

\author{J. Herbrych$^{1}$, P. Prelov\v sek$^{1,2}$ and X. Zotos$^3$}
\affiliation{$^1$J.\ Stefan Institute, SI-1000 Ljubljana, Slovenia}
\affiliation{$^2$Faculty of Mathematics and Physics, University of
Ljubljana, SI-1000 Ljubljana, Slovenia}
\affiliation{$^3$Department of Physics, University of Crete, and
Foundation for Research and Technology-Hellas, 71003 Heraklion, Greece}
\begin{abstract}
Finite-temperature Drude weight (spin stiffness) $D(T)$ is evaluated within  the anisotropic 
spin-$1/2$ Heisenberg model on a chain
using the exact diagonalization for small systems. It is shown that odd-side chains allow for more
reliable scaling and results, in particular if one takes into account corrections due to low-frequency 
finite-size anomalies.
At high $T$ and zero magnetization $D$ is shown to scale to zero approaching the
isotropic point $\Delta=1$.  On the other hand, for $\Delta>2$ at all magnetizations $D$ is 
nearly exhausted  with the overlap with the conserved energy current. Results for the $T$-variation
$D(T)$ are also presented.

\end{abstract}

\pacs{71.27.+a, 75.10.Pq}

\maketitle
\section{I. Introduction}
It has by now become evident  that many-body (MB) quantum systems of interacting particles
behave with respect to transport quite differently if they are either integrable or 
nonintegrable \cite{zprev,heid2}. In integrable systems the anomalous response shows up in a 
possibility of finite-temperature  
stiffness (Drude weight) $D(T)>0$ \cite{czp}, both the charge (or spin) and the
thermal one \cite{znp},   indicating the dissipationless d.c. transport at $T>0$. The prototype
model for this phenomenon is the anisotropic spin-$1/2$ Heisenberg  model on a 
chain, equivalent to the one-dimensional (1D) $t$-$V$ model of spinless fermions with 
nearest neighbor repulsion. Within this model one of the conserved quantities is the 
energy current $j_E$ leading  to the singular but trivial thermal dynamical-conductivity 
linear response \cite{znp},  i.e.,
$\kappa(\omega) = D_T \delta(\omega)$. On the other hand,  the spin current $j$ and the 
corresponding  dynamical spin conductivity (diffusivity) $\sigma(\omega)$ at $T>0$ is still the
subject of very active theoretical investigations and debate. 

In the case of a nonvanishing
projection of the spin current $j$ on local conserved quantities $Q_n$ the Mazur inequality offers a 
firm proof  of finite $D(T\neq 0) > 0$ \cite{znp} in the thermodynamic limit . Still, at zero magnetization,  
i.e., at the total spin  $S^z=0$ the overlap with all $Q_n$ vanishes independent of the anisotropy 
$\Delta$ \cite{znp}. To employ the same argument one possible path is to construct more general nonlocal
conserved quantities \cite{long,pros} which should be further explored. 

The (original) alternative formulation
via the MB level dynamics induced in a 1D ring via an external flux \cite{czp,sshsu} offers a qualitative
understanding and is the starting point for numerical  calculations
using the full exact diagonalization (ED) method \cite{zp,naro,heid1,rigo}. 
The latter so far did not eliminate
disagreement on several questions : a)  is $D$ a monotonous function of $\Delta$
at fixed $S^z$ \cite{heid1}, b) does $D(T>0) $ vanish on approaching the isotropic
point $\Delta=1, S^z=0$ \cite{naro,sirk}, c) which if any analytical result, obtained via 
the Thermodynamic Bethe Ansatz \cite{zot,benz}, is correct and compatible with numerical investigations. 

In the following we present results of the numerical study for $D(T)$ as obtained 
using the ED and the scaling for small systems. In contrast to previous works \cite{naro,heid1}
we perform the study within the canonical ensemble  which offers much faster convergence
with the chain size $L$, at least approaching the isotropic point $\Delta \sim 1, 
S^z \sim 0$. To avoid quite singular behavior of even-lengths chains, we study spin 
systems with odd $L$. In particular, we pay the attention to possible low-frequency 
contributions in the
dynamical conductivity $\sigma(\omega)$  which can give an insight into anomalies around 
commensurate $\Delta= \cos(\pi/\nu)$ with integer $\nu$, e.g.,
at $\Delta < 0.5$.

The paper is organized as follows: In Sec.~II we present the model and 
Drude weight $D$ as zero frequency contribution to dynamical 
conductivity. We shortly also describe numerical method used 
to analyse it. Our results are presented in Sec.~III. First we investigate 
the high-temperature limit $C=TD$, where we emphasize the low-frequency 
contributions which can mask the correct result. We show also that 
within Ising-type regime $\Delta>1$ the Drude weight calculated via the 
overlap with the conserved energy current gives nearly perfect results. 
Finally  we focus on the temperature variation of $D(T)$.

\section{II. Drude weight}
We study the anisotropic $S=1/2$ Heisenberg model on a chain with $L$ sites and periodic
boundary conditions
\begin{equation}
H=J\sum_{i=1}^{L} (S^x_iS^x_{i+1}+S^y_iS^y_{i+1}+\Delta S^z_iS^z_{i+1}), \label{xxz}
\end{equation}
where $S_i^\alpha$ are component of the $S=1/2$ spin operators. In order to define the 
Drude weight (spin stiffness) $D$ it is convenient to map the model (\ref{xxz}) via the Jordan-Wigner 
transformation  onto the $t$-$V$ model of interacting spinless fermions adding a fictitious 
magnetic flux  $\Phi = L \phi$ through the ring \cite{sshsu,kohn}, entering the  hopping matrix elements,
\begin{equation}
H=t\sum_i (e^{i\phi}c^\dagger_ic_{i+1}+ {\rm h.c.} )+
V\sum_i \Big(n_i-\frac{1}{2}\Big)\Big(n_{i+1}-\frac{1}{2}\Big), \label{tv}
\end{equation}
$n_i = c^\dagger_ic_i$, $t = J/2$ and $V= 2t \Delta$.  Here we consider only 
chains with odd number of fermions $N$ to avoid additional 
boundary fermionic sign and other finite-size effects discussed in more detail
below.  In the following we use everywhere 
$J=1$  in order to facilitate the comparison with the majority 
of previous works and references \cite{naro,zot,heid1}. Note that relevant 
parameters are now the total spin $S_z$ and magnetization $s=S_z/L$
or the fermion density or band filling $n=N/L=s+1/2$. 
 
Via the corresponding spin (particle or charge within the fermionic model) current
\begin{equation}
j= t\sum_i ( i e^{i\phi}c^\dagger_ic_{i+1}+ {\rm h.c.} ),
\end{equation}
one can express the dynamical (spin) conductivity at general temperature $T>0$ as
\begin{equation}
\sigma(\omega)=2\pi D\delta(\omega)+\sigma_{reg}(\omega),
\end{equation}
where the regular part $\sigma_{reg}(\omega)$ 
expressed in terms of eigenstates $|n\rangle$ and 
eigenenergies $\epsilon_n$, 
\begin{equation}
\sigma_{reg}(\omega)= \frac{\pi}{L}\frac{1-e^{-\beta\omega}}{\omega} \sum_{\epsilon_n\ne\epsilon_m}
p_n|\langle n|j|m \rangle |^2\delta(\epsilon_n-\epsilon_m-\omega), \label{sigreg}
\end{equation}
while the dissipationless component with the Drude weight (spin stiffness)
$D$ can be related to the flux dependence of MB states \cite{czp}, in analogy with the original
formulation by Kohn \cite{kohn}
\begin{equation}
D=\frac{1}{2L}\sum_n p_n\frac{\partial^2\epsilon_n(\phi)}{\partial\phi^2}, \label{stiff}
\end{equation}
where $p_n = \exp(-\beta \epsilon)/Z$ are corresponding Boltzmann factors.

The relation (\ref{stiff}) is convenient for the ED numerical evaluation of $D(T)$ in small systems, 
since it only  requires the calculation of eigenvalues $\epsilon_n(\phi)$. Finally we are
interested in the result within the thermodynamic limit $L \to \infty$ at fixed $T$ and magnetization $s$
(filling $n$) , hence several strategies to obtain the thermodynamic value
 are possible.  Since we mostly consider  the high-$T$ limit (allowing for most accurate ED results in small systems) 
and ED sizes are quite limited $L \leq 21$, we perform the canonical calculation at total spin $S^z$
(fermion number $N$).
The grand canonical evaluation  at available $L$ and high $T$ has a very broad distribution
of $N$, leading to overestimates of $D$ (or at least its slow convergence with $L$) in the 
vicinity of the isotropic  phase, i.e., at $s \sim 0 , \Delta \sim  1$.  On the other hand,
also results with even $L$ show deficiencies \cite{heid1}. Treating in Eq.(\ref{stiff}) the flux $\phi$ as
parameter, corresponding $D(\phi,T \gg 0)$ show strong anomaly at $\phi \to 0$ 
for even $L$ and even $N$ due to the particle-hole symmetry and degeneracy of MB levels.
In addition, even-$L$ systems give at odd $N$ considerably lower values 
for $D$ at $\Delta<1$ and small $L$ \cite{heid1} (an
origin could  be also particle-hole symmetry absent at odd $L$) remedied presumably only at 
much larger $L$. To avoid these complications, we in the following consider only systems 
with odd $L= 5 - 21$ (for $L=21$ only one $k$-vector due to very high CPU requirements) 
which reveal much weaker and  more regular $D(\phi)$  dependence.

\section{III. Results}
\subsection{A. High-Temperature Limit.}
In the following we mostly concentrate on the limit $T~\to~\infty$, expecting that obtained results
are quite generic and qualitatively similar at  any $T > 0 $. 
Since for $T\to \infty$, $D(T)$ scales as $1/T$
the relevant and nontrivial quantity is $C =T D(T)$, representing also the limiting value of the
current-current correlation function $C = C_{jj} (t \to \infty)$ \cite{znp}.  Let us first consider the most delicate 
zero-magnetization $s=0$ (half-filling $n=1/2$) case. Since we choose odd $L$, 
the actual calculations are performed 
for closest odd $N= (L \pm 1 )/2$.  Results for $C$ vs. $1/L$  for all odd $L=5 -19$ are presented 
for different $\Delta$  in Fig.~1. Several conclusions can be drawn directly from obtained results:
a) Both values as well as the scaling with $L$ are qualitatively different between $\Delta \geq 1$ and
$\Delta<1 $.  It is evident that for $\Delta \geq 1$ the only consistent
limit appears to be  $C = 0$. b) There are some visible anomalies near $\Delta <  0.5$
which indicate on a nonuniform dependence of $C(\Delta)$ \cite{heid1} and in particular 
different scaling $L \to \infty$ which we discuss in more detail below.

\begin{figure}
\includegraphics[width=0.47\textwidth]{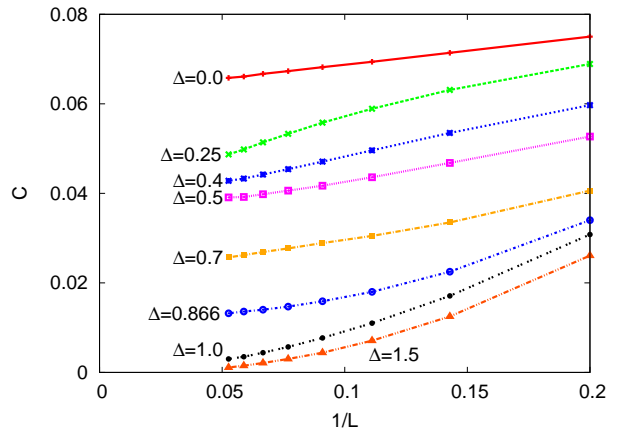}
\caption{(Color online) High-$T$ Drude weight  $C= T D$ vs. $1/L$  for 
zero magnetization $s=0$ and different $\Delta$ as obtained for systems with odd $L=5-19$. }
\label{fig1}
\end{figure}

In order to resolve the origin of the deviations of $C$ at $\Delta < 0.5$
as well as of quite regular convergence  of results for other values of $\Delta$
we investigate the dynamical $\sigma(\omega)$, shown conveniently also 
in the integrated form for $T \to \infty$,
 \begin{equation}
I(\omega)=C + \frac{T}{\pi}  \int\limits_0^\omega \sigma_{reg}(\omega')d\omega',
\end{equation}
consistent with the sum rule 
\begin{equation}
I(\omega \to \infty) = T e_{kin}=  - T  \langle H_{kin} \rangle  /L, \label{sumr}
\end{equation}
where $H_{kin}$ is the kinetic-energy part in the model 
(\ref{tv}).  $e_{kin}$  can be evaluated exactly in the $\beta \to 0$ 
limit, even for finite $L$ and fixed $N$, 
\begin{equation}
e_{kin} = \beta \frac{J^2}{4} \frac{N}{L}\Big(1-\frac{N-1}{L-1}\Big)
\end{equation}

\begin{figure}
\includegraphics[width=0.47\textwidth]{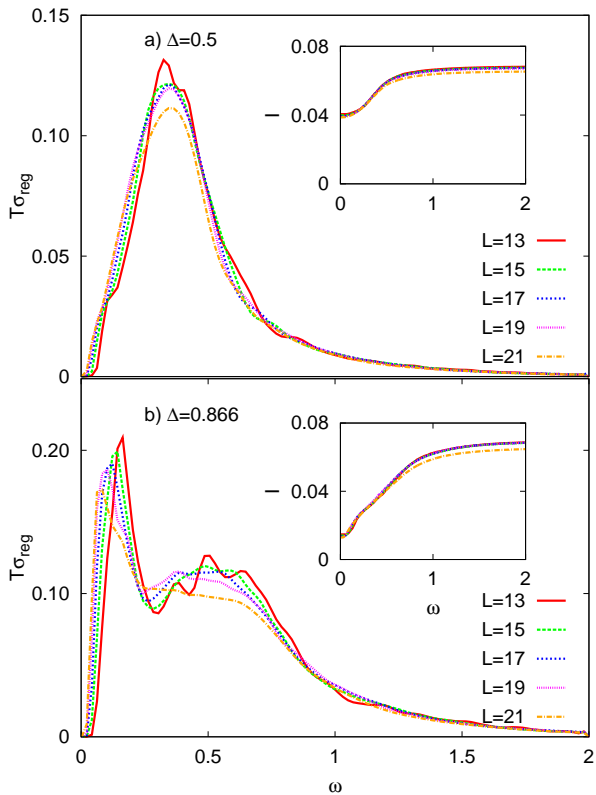}
\caption{(Color online) Regular part of dynamical conductivity 
$\sigma_{reg}(\omega)$ and the integrated
one $I(\omega)$ (inset) for $s=0$ and:  a) $\Delta=0.5$, and 
b) $\Delta=0.86 $ for different sizes $L=13- 21$.}
\label{fig2}
\end{figure}

In Fig.~2 we present characteristic results for $\sigma_{reg}(\omega)$ 
as well as $I(\omega)$  (in inset) for two commensurate values $\nu=3,6$, i.e., 
$\Delta=0.5, \sqrt{3}/2 = 0.866$, respectively. 
We note that for $\Delta=0.5$ the incoherent part in $\sigma_{reg}(\omega)$
is quite $L$-independent in a broad range $L=13-21$ and consequently the
convergence of obtained Drude weight $C$ vs. $1/L$  is very stable. Less obvious
case is $\Delta=0.866$ ($\nu=6$) being already closer to the critical value $\Delta=1$. 
The incoherent $\sigma_{reg}(\omega)$ reveals here a low-$\omega$ contribution 
whereby the peak is shifting as with $1/L$ as observed even
more pronounced for $\Delta>1$ \cite{elsha}. However, in the present case the
peak intensity as well diminishes with $L$ (a closer inspection reveals
that the peak $\omega_p$ also vanishes here faster than $1/L$)  so that the integrated 
$I(\omega)$ in Fig.~2b appears to have well defined limit $C=I(\omega \to 0)$.

In Fig.~3 we present $I(\omega)$ for $\Delta=0.25$ characteristic for the 
regime $\Delta<0.5$. We note that the  high-$\omega$ part is quite 
$L$-independent (note that for L=21 we calculate only one $k$-vector, 
which influences slightly the sum rule $I(\omega \to \infty)$) similar to results for $\Delta=0.5$ in  Fig.~2a. 
However, there is also a well visible anomalous low-$\omega$ contribution  
at $0.02< \omega < 0.08$  (see the inset). The peak in $\sigma(\omega)$ (as obtained from $I(\omega)$ in the 
inset of Fig.~3) appears
to shift towards $\omega_p = 0 $ somewhat faster than
$1/L$ (approximate fit $\omega_p \sim 1.342/L-0.017$) whereas its weight in $I(\omega)$
increases with the system size. 
This deviation can be counted  as an additional contribution to effective $\delta C$.
This  is, e.g.,  in contrast to case $\Delta=0.866$, where the intensity decreases with $L$ (Fig.~2).
Although the origin of the
low-$\omega$ anomaly is not well understood it seems that it is absent  for commensurate
values of $\Delta=\cos(\pi/\nu)$  which possess additional degeneracies \cite{zot}.
\begin{figure}
\includegraphics[width=0.47\textwidth]{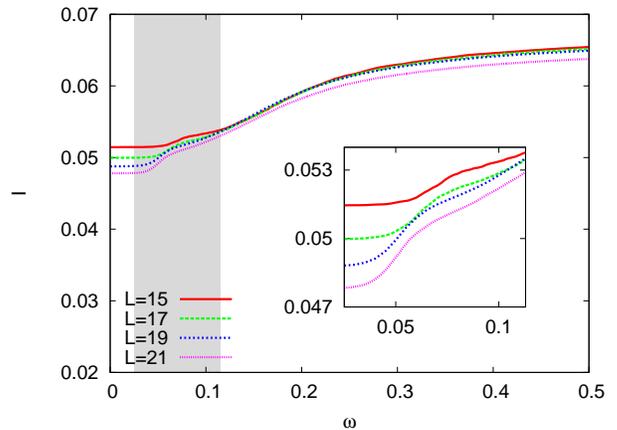}
\caption{(Color online) Integrated dynamical conductivity $I(\omega)$ for 
$\Delta =0.25, s=0$ and various sizes $L$. The inset focuses on 
the low-$\omega$ regime.}
\label{fig3}
\end{figure}

Results for $C$ vs. $1/L$ as in Fig.~1 can be used to extrapolated to
the thermodynamic value $C$ where we use the extrapolation
$C(L) = C + \alpha  / L + \zeta /L^2$. Obtained results for
$C(\Delta)$ are presented in Fig.~4. On the other hand, one can correct 
$C(L) $with the low-$\omega$ contribution $\tilde C(L)=C(L)  + \delta C(L)$
and get modified extrapolation $\tilde C$, also presented in Fig.~4.
We can now compare the results with the analytical result obtained 
via Thermodynamic Bethe Ansatz (TBA) \cite{zot,benz}, 
\begin{equation}
C =\frac{\gamma-\frac{1}{2}\sin(2\gamma)}{16\gamma},\quad
\Delta=\cos(\gamma), \label{ba}
\end{equation} 
the validity of which has been still questioned \cite{benz,sirk}. We note that the 
agreement  of the analytical form (\ref{ba}) with the corrected numerical 
$\tilde C$ is very satisfactory for $s=0$ within the whole regime of $\Delta$.

\begin{figure}
\includegraphics[width=0.47\textwidth]{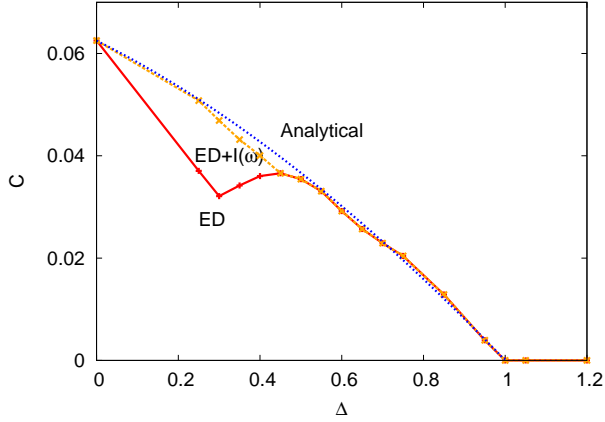}
\caption{(Color online) High-$T$ Drude weight $C=T D$ vs. $\Delta$ 
at magnetization $s=0$ obtained: 
using the ED and finite-size scaling of $D(L)$ (full curve), adding the low-$\omega$ 
correction (dashed curve), and within the analytical TBA \cite{benz,zot} (dotted curve).}
\label{fig4}
\end{figure}

Let us now turn to the dependence of $C$ on magnetization
$s$ (filling $n$). It is evident that one gets $C=0$ within the 
Ising-type regime  $\Delta>1$  only for $s=0$.  
Results for $C$ at $\Delta =1.7$ and $\Delta=3$
are shown in Fig.~5 for fixed $L=19$ and all available $S_z$.  It is indicative 
that  $C(s)$ are nearly equal for both $\Delta >1$. To go beyond the finite-size
results one can also perform the scaling to $L \to \infty$ analogous to $n=1/2$ case
which is possible, e.g., for $s=1/4$ (taking into account results for $L = 5- 25$) and 
$s=1/3$ (with results for $L=9 - 21$). Corresponding results for 
the extrapolated $C$ are also plotted in Fig.~5, confirming that $C(s)$ become
essentially universal for $\Delta>1$. 

It has been already observed in Ref.\cite{znp} that within the Ising regime $\Delta>1$ 
the Drude weight can be via the Mazur inequality well exhausted with the 
overlap onto the 
simplest nontrivial local conserved quantity $Q_3 = j_E$ representing the energy current. 
At $T \to \infty$ this overlap can be evaluated exactly leading to
\begin{equation}
C_3 = \frac{1}{2L}\frac{ \langle JQ_3\rangle ^2}{ \langle Q_3^2 \rangle} =
\frac{\Delta^2 s^2(1- 4 s^2)}{1+2\Delta^2(1 +4 s^2)}. \label{c3}
\end{equation}
From Fig.~5 we see that the agreement between the approximate $C_3$, Eq.(\ref{c3}), 
and the extrapolated $C$ is nearly perfect for large $\Delta \gg 1$, e.g., $\Delta=3$, while 
for $\Delta = 1.7$ the value $C_3$ starts to decrease, so that  $C_3<C$. 
In fact we observe from Eq.(\ref{c3}) that $C_3$ just saturates as a function of
$\Delta$ for  $\Delta \gtrsim 1.7-2$ and its value there  can already reasonably reproduce $C$.
We stress again completely different behavior is for $\Delta<1$ and $s=0$. 
In this case one gets $C_3=0$ (as well as higher overlaps  $C_{n>3}=0$ due to particle-hole symmetry), 
hence the Mazur inequality with local conserved quantities is unable to reproduce $C>0$ 
at $s=0$ \cite{znp}.

\begin{figure}
\includegraphics[width=0.47\textwidth]{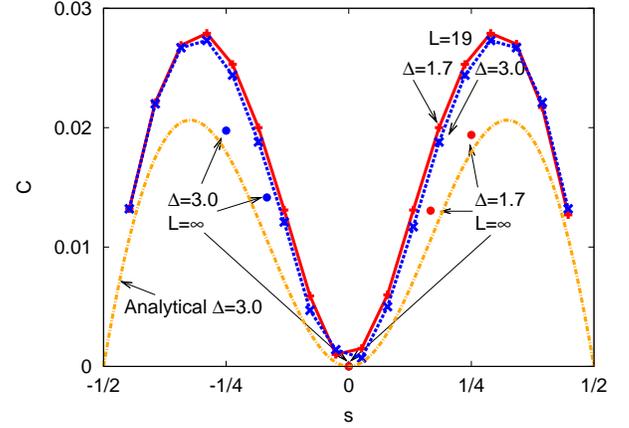}
\caption{(Color online) High-$T$ Drude weight $C$ vs. magnetization $s$ within the Ising-type
regime $\Delta = 1.7, 3.$ obtained for fixed $L=19$, with the $1/L \to 0$ extrapolation 
for $s=0, 1/4, 1/3$ (dots), and the analytical approximation $C_3$, Eq.(\ref{c3}), 
for $\Delta= 3$.} \label{fig5}
\end{figure}

Let us further consider the normalized Drude weight $D^*=D/ e_{kin}$ which
represents the relative weight of the dissipationless transport within the whole sum rule,
Eq.(\ref{sumr}), i.e., we have $0<D^*<1$. 
Since one cannot perform a systematic extrapolation $L \to \infty$ for arbitrary magnetization
$s$ we present in Fig.~6 results for $D^*$ within the whole (half) plane 
$\Delta,  s \geq 0$ as calculated in systems with fixed $L=19$. Apart from some anomalies 
observed  (without the correction $\delta C$) already in Fig.~3 we confirm quite regular 
dependence $D^*$ on $(\Delta, s)$. It is
quite evident that in the limiting case $\Delta =0$ (XY model)  we get 
$D^*=1$ corresponding to noninteracting fermions where the whole sum rule is 
within the Drude weight. The same hold for maximal magnetization $s \to \pm 1/2$ 
(for nearly empty or full band, $n \to 0, n \to 1$, respectively) 
where the interaction does not play a role. 
For fixed $\Delta$ the minimum of $D^*$ is always at
$s = 0$ whereby the dependence $D^*(s)$ is nearly universal
for all $\Delta >1$.    

\begin{figure}[floatfix]
\includegraphics[width=0.47\textwidth]{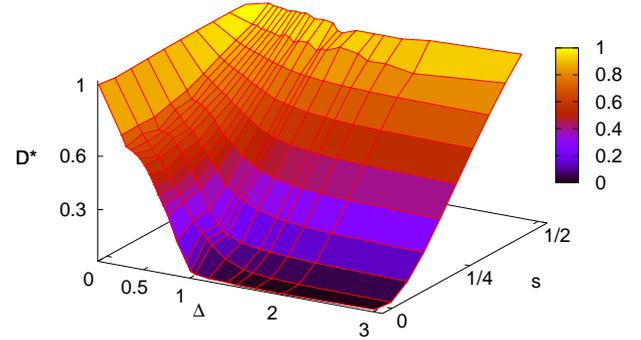}
\caption{(Color online) Normalized Drude weight $D^*$ within the plane 
$ \Delta, s$ as calculated in systems with fixed size $L=19$. }
\label{fig6}
\end{figure}

\subsection{B. Finite Temperature}
Finally, let us present results for the $T$-dependence $D(T)$ as evaluated 
using the relation (\ref{stiff}), again restricting our analysis to zero magnetization $s=0$
and systems with odd $L$ (Fig.~7). It should be realized that numerical results at low $T<0.5$ 
are more susceptible to finite-size effects since very small number of MB levels
effectively participate in $D(T)$ and the crucial contribution comes from the 
ground state $\epsilon_0(\phi)$. Still, in spite of some discrepancies at low $T<0.4$
the overall agreement  with the TBA result \cite{zot} is reasonable. Another conclusion is that
the extended high-$T$ behavior, i.e. $D=C/T$ is followed very accurately down to
quite low $T > 0.5$ in the whole range $\Delta <1$.  While the ground state value $D_0$ 
is quite reliable in the intermediate window $0<T<0.5$ results are
sensitive to finite-size effects so we cannot give a firm conclusion on possible 
nonanalytical low-$T$ behavior as predicted in Ref.\cite{zot}.

\begin{figure}
\includegraphics[width=0.43\textwidth]{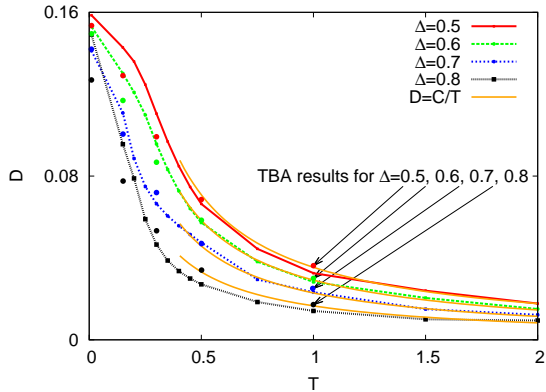}
\caption{(Color online) Finite-$T$ Drude weight $D(T)$ for $\Delta<1$ at magnetization $s=0$
as calculated numerically 	
using ED with $L=15-21$ and finite-size scaling (full line with dots), extrapolating the high-$T$
numerical result, i.e. $D=C/T$ (thin lines), and Thermodynamic Bethe Ansatz result (dots) 
from Ref.\cite{zot}.}
\label{fig7}
\end{figure}

\section{IV. Conclusions}
In conclusion, we have shown that numerical evaluation of the Drude weight (spin stiffness)
$D(T)$ within the anisotropic Heisenberg model 
can lead to more controlled and converged results if performed in
a canonical ensemble, at fixed $S_z$ (number of particles $N$).
Breaking of the particle-hole symmetry by using systems with odd $L$ is also 
helpful and is advantageous over usually studied systems with even $L$.
Our study is mostly concentrated on the high-$T$ limit which should be anyhow
quite generic for the whole regime $T>0$. 
Results obtained at zero magnetization $s=0$  using the finite-size 
scaling confirm the change of character of $D(T)$ at $\Delta=1$, i.e., they are compatible
with the $D(T)=0$ for $\Delta>1$. While at $s=0$ within the majority of the regime 
$\Delta<1$ there are no evident problems with the scaling $1/L$ of $D(T)$ we
have traced the irregularities at $\Delta <0.5$ back to the emergence of finite-size
low-$\omega$ contribution in $\sigma_{reg}(\omega)$ which can lead to a
finite correction $\delta C$ in the thermodynamic limit $L \to \infty$.
Taken the latter into account, we find a very good agreement with the TBA result
\cite{zot}, in this way possibly eliminating (or at least restricting) some recently 
expressed questions regarding its validity. 

High-$T$ normalized Drude weight $D^*$ away from $s=0$ 
shows a systematic and smooth variation with $s$ towards the limiting values
$D^*=1$ for $s=\pm 1/2 $ as well as in XY limit $\Delta=0$. In the Ising regime 
$\Delta>1$ (in particular for large $\Delta>2$) the variation $C=TD(s)$ is very well 
reproduced with the Mazur inequality overlap with the conserved energy current $j_E$,
in very contrast to the XY-type regime $\Delta<1$.

Results for the $T$-variation $D(T)$ reveals that even quantitatively 
the high-$T$ result $D=C/T$ remains valid in a wide regime, i.e., generally
for $T>0.5$. While small-system results allow also for a reliable scaling for 
$D_0=D(T=0)$ at $s=0$,  the finite-size effects are rather hard to avoid
in the window $0<T<0.5$ and other methods  beyond the ED are needed to
investigate in more detail this regime.

The authors acknowledge the support of the RTN-LOTHERM project and
the Slovenian Agency grant No. P1-0044.

\end{document}